\documentclass[conference]{IEEEtran}
\IEEEoverridecommandlockouts
\usepackage{cite}
\usepackage{amsmath,amssymb,amsfonts}
\usepackage{algorithmic}
\usepackage{graphicx}
\usepackage{textcomp}
\usepackage{xcolor}
\usepackage{hyperref}
\usepackage{listings}
\newcommand{\blackwell}{\textcolor{blue}{\textbf{GB203}}}

\hypersetup{
    colorlinks=true,  
    linkcolor=blue,   
    filecolor=magenta, 
    urlcolor=red,     
    citecolor=red
}
\usepackage{url}
\usepackage{array}
\usepackage{enumitem,kantlipsum}
\usepackage{float}

\def\BibTeX{{\rm B\kern-.05em{\sc i\kern-.025em b}\kern-.08em
    T\kern-.1667em\lower.7ex\hbox{E}\kern-.125emX}}
\begin{document}

\title{Dissecting the NVIDIA Blackwell Architecture with Microbenchmarks\\
}
\author{
\IEEEauthorblockN{Aaron Jarmusch}
\IEEEauthorblockA{\textit{dept. of Computer Information Sciences} \\
\textit{University of Delaware}\\
Newark, US \\
ORCID}
\and
\IEEEauthorblockN{Nathan Graddon}
\IEEEauthorblockA{\textit{dept. of Computer Information Sciences)} \\
\textit{University of Delaware}\\
Newark, US}
\and
\IEEEauthorblockN{Sunita Chandrasekaran}
\IEEEauthorblockA{\textit{dept. of Computer Information Sciences)} \\
\textit{University of Delaware}\\
Newark, US \\
schandra@udel.edu or ORCID}
}

\maketitle

\begin{abstract}
\label{abstract}

The rapid development in scientific research provides a need 
for more compute power, which is partly being solved by GPUs. 
This paper presents a microarchitectural analysis of the modern
 NVIDIA Blackwell architecture by studying GPU performance 
 features with thought through microbenchmarks. We unveil key 
 subsystems, including the memory hierarchy, SM execution 
 pipelines, and the SM sub-core units, including the 5th 
 generation tensor cores supporting FP4 and FP6 precisions. 
 To understand the different key features of the NVIDIA GPU, 
 we study latency, throughput, cache behavior, and scheduling 
 details, revealing subtle tuning metrics in the design of 
 Blackwell. To develop a comprehensive analysis, we compare the 
 Blackwell architecture with the previous Hopper architecture by 
 using the GeForce RTX 5080 and H100 PCIe, respectively. We 
 evaluate and compare results, presenting both generational 
 improvements and performance regressions. Additionally, we 
 investigate the role of power efficiency and energy consumption 
 under varied workloads. Our findings provide actionable insights 
 for application developers, compiler writers, and performance 
 engineers to optimize workloads on Blackwell-based platforms, 
 and contribute new data to the growing research on GPU 
 architectures.

\end{abstract}


\section{Introduction}
\label{sec:introduction}

With recent and rapid advancements in Artificial Intelligence 
(AI), GPUs have become a compelling resource for accelerating 
machine learning and high-performance computing (HPC) workloads. 
Their massively parallel architecture makes them well-suited 
for a wide range of applications spanning several domains.
Offering significant improvements in computing capabilities 
for applications that once required days, or even years, of 
computation can now be solved in a matter of hours or minutes. 
This shift underscores the growing demand for powerful, 
efficient GPUs in both scientific and industrial research. 

Over the past decade, these hardware accelerators have become 
increasingly competitive. Vendors such as NVIDIA, AMD, Intel, 
and Google have each introduced specialized accelerators, from 
Tensor Processing Units (TPUs) to customized GPUs, conformed 
to specific computational needs. With such a wide range of 
options, a natural question becomes apparent. How do we 
determine which architecture best suits a given workload?

Many researchers have developed a variety of tools and 
methodologies to provide a solution. Application 
profiling\cite{5336219}, 
roofline models\cite{leinhauser2021metricsdesigninstructionroofline}, 
analytical performance modeling\cite{Hong2009}, 
and cache stall predication\cite{wenhao2012}, to list a few, 
have all been used to provide insights into these accelerators 
to gain the best performance for specific applications. 
Though, another approach involves dissecting GPU architectures 
at the microarchitectural level to identify compute-dependent 
features. However, this strategy is often hindered by the lack 
of public documentation on modern commercial GPUs, which limits 
the depth of a thorough analysis. 

Academic studies have studied the inner workings of NVIDIA's 
earlier architectures~\cite{Wong2010,Subramoniapillai2012,jia2018,jia2019,luo2025}. 
NVIDIA's Hopper (chip name \textbf{GH100})\cite{NVIDIA2022_H100} 
and Blackwell (chip name \blackwell{})\cite{NVIDIA2024_Blackwell} 
architectures represent two ends of their design. 
The \textbf{GH100} is optimized for large-scale AI training 
and scientific simulations\cite{fusco2024understandingdatamovementtightly}, 
while the \blackwell{} targets real-time graphics and inference 
workloads in power-constrained environments\cite{NVIDIA_RTX_Blackwell}. 
Although these GPUs share a similar layout, they diverge 
significantly in terms of hardware configuration and 
memory hierarchy. 

This paper presents a microarchitectural comparison of the 
\textbf{GH100} and \blackwell{} GPUs with a basis of 
real-world performance subsystems, such as shared memory, 
L1 and L2 cache behavior, core execution pipelines, and 
tensor core instruction throughput. Our work presents a 
cohesive analysis with designed microbenchmarks in PTX and 
CUDA to reveal subtle yet important differences in the 
behavior of these architectures under stress, especially in 
compute-bound and memory-bound applications. 

A unique aspect of this comparison is revealed in the 
intended purposes of these GPUs. The \textbf{GH100} equipped 
with HBM2e memory and high SM count, tailored for maximum 
throughput and data locality in AI training. On the other hand, 
the \blackwell{} trades cache size and double-precision 
capability to achieve higher clock rates and efficiency 
within a consumer-friendly GPU. By analyzing the response 
of each architecture to instruction-level parallelism, 
memory coalescing patterns, and warp scheduling pressure, 
we aim to provide insight into the core principles of 
scalability and specialization that underscores NVIDIA's 
Blackwell architecture for practical insights for application 
developers, performance engineers, and compiler creators.

The key contributions of this work are as follows:
\begin{itemize}
    \item Develop a new set of microbenchmarks to evaluate key components of the NVIDIA Blackwell architecture, with a comparison to the previous Hopper generation GPUs. 
    \item Perform an in-depth analysis of the memory hierarchy and Streaming Multiprocessor (SM) subunits, including the next-generation tensor cores, unified INT32/FP32 cores, and FP64 execution cores.
    \item Provide performance guidelines to software as well as application developers such that they  effectively tap into the rich benefits of the hardware. 
    \item Explore the behavior and performance implications of newly supported low-precision floating-point datatypes, FP4 and FP6, within Blackwell's tensor cores. 
\end{itemize}

We are unable to share the code at this time due to the blind-review policy, but we plan to open-source it post the review process and the outcome.

\section{Related Work}
\label{sec:related-work}
Understanding GPU performance has long been a critical focus in 
HPC research. Over the years, several studies have used 
microbenchmarking and other types of benchmarking to observe 
the abstract layers and analyze GPU architectures in fine gained details. 
Early works, such as \cite{Wong2010, Subramoniapillai2012}, 
focused on the first generations of GPU architectures like 
NVIDIA's Tesla and Fermi. These work focused on profiling 
memory access patterns and evaluating cache hierarchies. Their 
work laid the foundation for many of the benchmarking strategies 
still in use today. 

As newer architectures emerged, analyses 
targeted Kelper \cite{Zhang2017}, Maxwell\cite{Mei2012}, and Pascal,
 often focusing on warp scheduling, instruction latency and memory 
 coalescing behavior. With the release of the Turning, Volta, 
 Ampere, and Hopper architectures, studies \cite{Fasi2021, jia2019, tan2011, Markidis2018, Martineau2018, Raihan2019, Yan2020, luo2025} 
shifted toward evaluating mixed-precision units and tensor core 
performance. These work introduced microbenchmarks that probed 
\(mma\) instruction latencies, tile sizes, and data layouts. 
Some also addressed instruction-level parallelism (ILP) \cite{Sun2022}, 
examining how GPU pipelines behave under high register pressure 
or deeply nested loops. In parallel, Hong and Kim \cite{Hong2009} 
developed an analytical model for GPU architectures. This laid 
the ground work for tools such as Accel-Sim\cite{accelsim2020} and 
GCoM \cite{gcom2022} to develop an implementation for modern GPUs 
by simulating architectural behavior. While powerful, these 
tools often require detailed hardware understanding or lack 
support for new instructions introduced in newer architectures 
like Blackwell. 
Even though lately, it seems Large Language Models (LLMs) can be used 
to simulate code execution on GPUs, the understanding of 
microarchitectures is still necessary work to improve
simulating code execution on GPUs \cite{Khoi2025}. 
Despite these existing works, little has been published on the 
architectural features specific to Blackwell. Our work aims to 
fill this gap. 

To the best of our knowledge our work provides 
the first detailed microbenchmark based study of Blackwell's 
core subsystem, including FP64 execution, low-precision MMA 
units, and memory throughput with shared, L1, and L2 caches. 

\textit{By building on existing work and extending benchmarking methodologies to Blackwell's novel features, we contribute new insights to the evolving landscape of GPU performance analysis. }

\section{Overview of Blackwell Architecture}
\label{sec:overview}
Although the \blackwell{} is based off the \textbf{GH100}, 
they represent two design philosophies. \textbf{GH100} - NVIDIA's Hopper architecture GPU chip - is optimized for 
 large-scale AI and scientific computing workloads, while the 
 \blackwell{}, part of the Blackwell architecture, is a 
 power-efficient, consumer-focused GPU design to support gaming, 
 rendering, and small-batch inference tasks. Both implement a 
 similar CUDA core programming model with differences in 
 instruction sets, execution unit counts, memory hierarchy, 
 and resource scheduling.

\subsubsection{\textbf{SM and Execution Pipeline}}
At the center of both architectures is the Streaming 
Multiprocessor (SM), which is responsible for warp scheduling, 
instruction issue, and execution. While the high-level SM 
design remains similar there are several differences shown 
in Table \ref{tab:sm-config}.
\begin{table}
    \centering
    \scriptsize
    \begin{tabular}{|c|c|c|}\hline
         &  \textbf{GH100}& \blackwell{}\\\hline
 Architecture& Hopper&Blackwell\\\hline
 GPU Name& H100 PCIe&RTX GeForce 5080\\\hline
         FP32 per SM&  128& Unified INT32/FP32 Unit\\\hline
         INT32 per SM&  64& Unified INT32/FP32 Unit\\\hline
         FP64 per SM&  64& 2\\\hline
         tensor cores&  4th gen& 5th gen (supports FP4 and FP6)\\\hline
         Transformer Engine&  1st gen& 2nd gen\\ \hline
    \end{tabular}
    \vspace{.2cm}
    \caption{The execution units on the \textbf{GH100} and \blackwell{} GPUs.}
    \label{tab:sm-config}
    \vspace{-0.5cm}
\end{table}
Blackwell introduces improvements in warp scheduling, 
reducing warp dispatch latency for divergent workloads. 
\textbf{GH100}, on the other hand, offers greater execution 
throughput and larger on-chip buffering for training-scale 
workloads.


\subsubsection{\textbf{Cache Hierarchy}}

\begin{table}[t]
    \centering
    \scriptsize
    \begin{tabular}{|l|l|l|}\hline
         Memory Unit&  \textbf{GH100}& \blackwell{}\\\hline
         L0 i-cache& separate partition & unified\\\hline
         Register File Size (KB/SM)& 256 &256 \\\hline
         L1 cache Size (KB/SM)&  256 (unified)& 128 (unified)\\\hline
         Shared Memory Size (KB/SM)&  228 (unified)&  (unified)\\\hline
         L2 Cache (Size MB/SM)&  50 (2 partitions)& 65 (1 partition)\\\hline
         Global Memory (GB)&  80 (HBM2e)& 16 (GDDR7)\\ \hline
    \end{tabular}
    \vspace{.2cm}
    \caption{The cache hierarchy and partition sizes of the \textbf{GH100} and \blackwell{}.}
    \label{tab:cache-hierarchy-details}
    \vspace{-0.5cm}
\end{table}

A significant difference between the two GPUs is in the cache 
and memory subsystem. Table \ref{tab:cache-hierarchy-details} 
shows a detailed comparison and \blackwell{} compensates 
for its smaller L1 with increased L2 bandwidth. \textbf{GH100}'s 
HBM enables larger batch sizes and working set capacity, helpful 
for training large models.

\subsubsection{\textbf{Tensor and AI Acceleration}}
Tensor cores are specialized units designed to accelerate 
matrix operations. Hopper introduces the 4th generation 
tensor cores with FP8 workload support, and Blackwell 
advances to the 5th generation with extended support for 
FP4/FP6 formats while maintaining FP8 workload support.

\subsubsection{\textbf{Instruction Set and Software Compatibility}}

With the Hopper architecture, support for tensor core 
instructions like \(wgmma\) and FP8 arithmetic are via 
PTX and CUDA 11.8. They maintain backward compatibility with 
traditional \(mma\) and CUDA C++ instructions.
CUDA 12.8 and PTX 8.7 expands support for Blackwell's 5th-gen tensor core 
instructions, including \(tcgen05\) and enhanced operand types 
with formats for FP4 and FP6. Though the Hopper \(wgmma\) instructions 
are not compatible with Blackwell, instead the new \(tcgen05\) 
instructions can be used for warp-group computation. 

Together, \textbf{GH100} is a training optimized architecture 
with expansive memory resources, while the \blackwell{} 
focuses on maximizing energy efficiency within thermal 
constraints. Despite their differences, both share 
microarchitectural similarities for a shared benchmarking 
strategy. 


\textit{Subsequent sections of this paper dissect each of 
these features through targeted microbenchmarks and analysis 
from the Hopper H100 PCIe (\textbf{GH100} chip) and Blackwell GeForce 
RTX 5080 (\blackwell{} chip) GPUs. The paper analyzes Blackwell at depth but also demonstrates a comparison between the previous 
 architecture.}

\section{Compute Pipeline}
\label{sec:sm-cores}
Most computation on the GPU is done by a compute 
pipeline and the center of this pipeline is the SM. The SM is responsible 
for allocating compute resources, scheduling 
instructions, and moving data. To complete this task, 
each SM includes four partitions of execution units 
or sub-cores for integer, floating-point and 
tensor operations. The arrangement and hardware 
implementation of these sub-cores can provide 
improvements or impede application performance. 
Understanding these resources can enable developers 
to improve their applications while fully utilizing 
their GPU. To offer a comprehensive evaluation on 
the \blackwell{} and \textbf{GH100}, here are 
some metrics and analysis:

\begin{enumerate}[leftmargin=*]
    \item \textbf{Latency}: 
    Referring to the number of clock cycles an 
    instruction takes to produce a result that is 
    usable by a subsequent instruction. We measure 
    \textbf{True latency}, which is a serialized, 
    dependent instruction chain. Reflecting 
    the data-ready delay without parallelism or overlap.
     When there is a set of independent instructions 
     that are allowed to overlap and parallelize this 
     is called \textbf{Completion latency}. Both 
     forms of latency are reported in clock cycles 
     per instruction \((\#clock\_cycles/\#instruction)\).
     
    \item \textbf{Throughput/Bandwidth}: 
    Measured by the number of 
    instructions completed per clock cycle per SM, 
    based on measured runtime and instruction count.  
\end{enumerate}
We wrote our microbenchmarks in PTX instructions and CUDA. Instead of writing inline instructions within CUDA files we wrote  PTX kernels within the individual files, that will be executed with a compiled CUDA file at runtime.  These separate PTX kernel files prevents compile
time optimizations to the code. To confirm the PTX instructions weren’t optimized during runtime we review the generated SASS code from each PTX kernel file, to confirm there have been no optimizations to the instructions.
\subsection{Clock Overhead}
All clock cycles are measured with \(\%clock64\), a 
predefined read-only special register that returns 
the counter value of the clock cycle when read\cite{NVIDIA_PTX_8.8}. 
For example, Listing \ref{lst:clock} takes the value of 
the counter before and after the instruction, then 
subtract the values for the measured clock cycles. 

\begin{figure}[H]
\begin{lstlisting}
    .reg .u64 %start, %end;
    ld.param.u64 %time_ptr, [param_time];
    mov.u64 %start, %clock64;
    mad.lo.s32 r1, r1, r2, r3;
    mov.u64 %end, %clock64;
\end{lstlisting}
\caption{PTX code measures the clock cycles of a \texttt{mad.lo.s32} instruction.}
\label{lst:clock}
\end{figure}

On \blackwell{}, without an instruction between 
the registers, the subtracted value is 1, compared to 
a value of 2 for \textbf{GH100}. Additionally, when 
wrapping a combination of instructions, i.e. a mixed workload, 
the value is dependent on the instructions, which can
 shed insight into instruction workflows. 

\subsection{INT and FP32 Execution Units}

In previous architectures like Volta and Ampere, 
INT32 and FP32 instructions were issued through 
separate execution pipelines, often leading to 
suboptimal utilization when workloads were 
dominated by one type\cite{jia2018}. 

In \blackwell{}, NVIDIA includes unified execution 
units that can handle both INT32 and FP32 instructions,
 enabling dynamic scheduling based on instruction mix. 
 Potentially reducing idle cycles and improving 
 throughput for mixed workloads.

However, the unified cores can only be used as 
INT32 or FP32 operations during any clock cycle. 
Potentially, creating a hazard during mixed INT32/FP32 
workloads.

To accurately test the functionality of these 
unified cores, we utilized standard arithmetic 
and integer operations using the PTX instructions, 
\(fma\) and \(mad\), for FP32 and INT32, respectively. 
 We compare results from three sets of kernels, 
 (A) Pure INT32, (B) Pure FP32, and (C) Mixed 
 INT32/FP32 instructions to represent mixed 
 workloads utilizing both operations. Each workload was 
 executed 1024 times and averaged to provide a 
 noise-less understanding on these cores. 



\begin{table}[]
    \centering
    \scriptsize
    \begin{tabular}{|c|c|c|c|c|c|}\hline
         GPUs &  Pure INT32&  Pure FP32& Mixed 1 & Mixed 2& Pure FP64\\\hline
         GB203&  4/16.97&  4/7.97& 15.96/14 & 26.28/18 & 63.57/11 \\\hline
         GH100&  4/16.69& 4/7.86& 31.62/16 & 43.54/20 &  8.04/13 \\\hline
    \end{tabular}
    \vspace{.2cm}
    \caption{\textbf{GH100} vs \textbf{GH203} latency results (true/completion).}
    \label{tab:int32fp32_latency}
    \vspace{-0.5cm}
\end{table}

Our measurements show the true latency for pure 
INT32 and FP32 workloads on both GPUs were four cycles.
 \textbf{GH100} did slightly better on the completion 
 latency with pure kernels. Although \textbf{GH100} 
 uses separate execution pipelines for INT32 and FP32 
 operations, we observed \textbf{GH100} performs worse than the 
 \blackwell{} when executing the mixed instruction 
 sequence, see Table \ref{tab:int32fp32_latency}. 
 This suggests the unified INT32/FP32 execution 
 cores on Blackwell introduced more efficient 
 compute pipelines. 
To mention, \blackwell{} had a higher true and 
completion latency for the pure and mixed kernels
during the first run, which is absent from the 
results in Table \ref{tab:int32fp32_latency}. Other 
studies have also excluded similar 
results \cite{6844867} due to missed cache access 
before the cache warmed up. Though, this was not 
present with the Hopper run.

\textit{The Blackwell architecture shows improvement in latency in the mixed workloads while Hopper does better with the pure instruction workloads.
}
\subsection{FP64 Execution Units}

Double-precision (FP64) execution units are essential 
for workloads requiring high numerical accuracy, 
such as scientific simulations. While AI and graphics 
applications increasingly rely on low-precision formats
 (FP16, BF16, FP8, etc.), FP64 remains a key feature 
 for HPC and research domains.

\textbf{GH100} and \blackwell{} each contain a 
dedicated set of FP64 execution units, 
physically separate from the INT32/FP32 units. 
This separation allows the scheduler to issue 
FP64 instructions independently. The \blackwell{} 
chip has two FP64 execution units per SM, compared to 
\textbf{GH100} which has 64, see 
Table \ref{tab:sm-config}. 

Our microbenchmarks displayed expected results, 
Table \ref{tab:int32fp32_latency} shows the \textbf{GH100}
 latency for 1024 FP64 dependent instructions remained 
 below the \blackwell{} When only two dependent 
 instructions were executed, the \blackwell{} latency 
 decreased to 37.5 cycles. Normally, running more 
 instructions hides latency, however, in this case 
 there are only two execution units. This suggests the 
 two units are only for type and instruction support
  while the calculation is meant to be emulated with 
  other precisions i.e. using the FP32 executions units 
  or the tensor core instead. 

\textit{These insights are particularly relevant for users targeting portable performance across both datacenter and consumer-grade GPUs, where understanding FP64 bottlenecks can inform precision tradeoffs or algorithm design.
}
\subsection{Warp Scheduler Behavior and Issue Model}

\begin{figure}
    \centering    \includegraphics[width=1\linewidth,height=4cm,keepaspectratio]{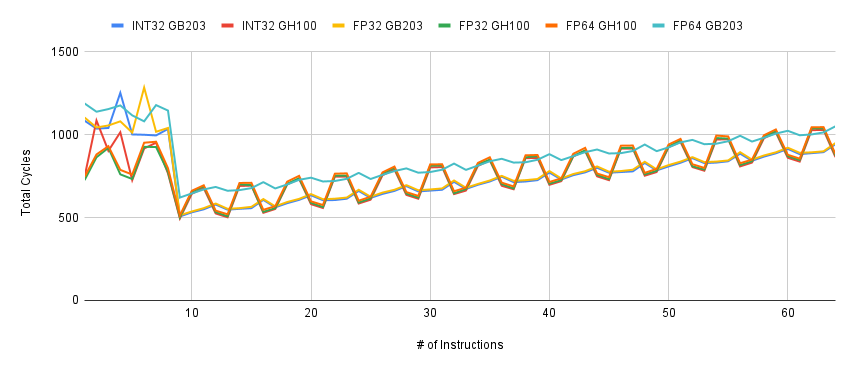}
    \caption{Comparing Total Cycles vs Iterations of the \blackwell{} and \textbf{GH100} GPUs with INT32, FP32, and FP64 workloads. }
\label{fig:intfp32_tcycles}
    \vspace{-0.3cm}
\end{figure}

\begin{figure}
    \centering    \includegraphics[width=1\linewidth,height=4cm,keepaspectratio]{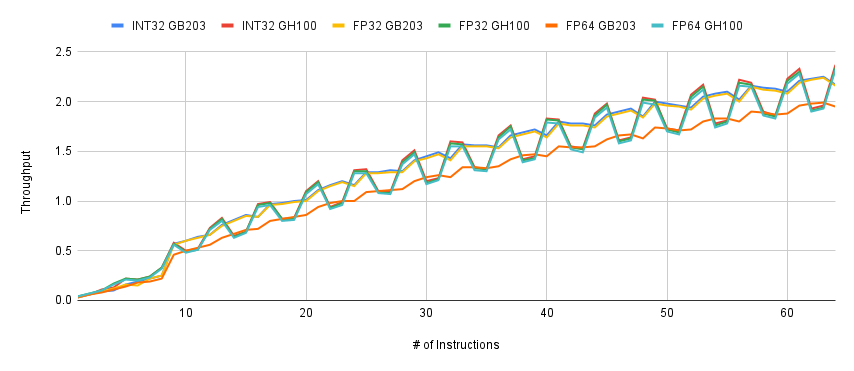}
    \caption{Comparing Throughput vs Iterations of the \blackwell{} and \textbf{GH100} GPUs with INT32, FP32, and FP64 workloads. }
    \label{fig:intfp32_throughput}
    \vspace{-0.5cm}
\end{figure}

To evaluate warp scheduling sensitivity and latency 
handling under dependency chains, we implemented a 
serialized dependent-instruction benchmark, where each thread executes a chain of dependent arithmetic operations in registers for each execution unit.
Figures \ref{fig:intfp32_tcycles} and 
\ref{fig:intfp32_throughput} show results as we incremented the number of dependent instructions per thread from 1 to 1024 (results shown for 1 to 64). To ensure a fair comparison of total cycles and throughput, we control the total number of instructions executed by adjusting the loop iteration count for each chain length.

As shown in Figure \ref{fig:intfp32_throughput}, 
throughput increases steadily for the first 1-9 
instructions, after which the pipeline differs 
between architectures and incrementally improves 
for all INT32, FP32, and FP64 workloads. The low throughput for short dependent chains is due to insufficient ILP within each thread, which limits the GPU’s ability to hide instruction latency. With only a few dependent instructions, threads quickly stall, and the scheduler cannot fully utilize the execution units, resulting in lower performance. As the chain length increases, the scheduler can better overlap execution and hide latency, improving throughput.

Notably, \blackwell{} exhibited smoother, more consistent increase 
in throughput compared to \textbf{GH100}'s more irregular 
ramp-up. Except, \blackwell{} demonstrated a lower 
throughput with the FP64 instructions compared to 
the INT32 or FP32 counterparts.

Similarly, Figure \ref{fig:intfp32_tcycles} shows 
\textbf{GH100} achieves lower total cycles before 
8 instructions, indicating more effective latency 
hiding under short dependency chains, compared to 
\blackwell{}. After 8 instructions the total 
cycle count sharply drops for both architectures, 
likely due to instruction scheduling warming up or 
pipeline effects. Though, similar to throughput, 
\textbf{GH100} has more sporadic total 
cycles as the instruction count increases.

The slightly better throughput from the \textbf{GH100} 
in small number of dependent instructions is likely 
due to deeper instruction buffering and a more 
aggressive warp scheduler that can tolerate 
instruction dependencies when pressure is high. 
However, aggressive scheduling can also introduce 
instability during the higher number of instructions. 
On the other hand, \blackwell{}'s steadier 
progression suggests a more conservative issue 
strategy. However, the higher total cycles during 
small number of dependent instructions does need 
more analysis. 

\textit{While both GPUs exhibit comparable performance overall, \textbf{GH100} tolerates short latency bound instruction sequences, whereas \blackwell{} is optimized for more regular high-ILP kernels.
}


\section{5th Generation Tensor Core}
\label{sec:tensor-core}

Introduced in the Volta architecture, tensor cores are 
specialized units designed to accelerate matrix 
multiplication, a foundational operation in deep 
learning and scientific computing. To fully utilize 
these tensor, kernels with matrix multiplication 
operations are used, such as matrix multiplication 
accumulate (MMA). To observe the behavior of the 
\blackwell{} and \textbf{GH100} our custom PTX 
microbenchmarks measure execution latency, throughput, 
and operand staging behavior. This section evaluates 
the \blackwell{} and \textbf{GH100}'s tensor cores 
capabilities ILP and varying warp counts. 

\subsection{Instruction Sets and Supported Datatypes}

The fifth-gen \blackwell{} and fourth-gen \textbf{GH100} 
tensor cores both support instructions with different 
datatypes, operand handling, and performance tuning. 
Table~\ref{tab:tensor-core-data-types} compares the 
supported instructions and datatype precisions between 
the fourth and fifth generations of the tensor core. 

The fifth generation in Blackwell introduces new datatypes (FP4 and FP6) implemented in CUDA with 
new SASS-level instructions (e.g., OMMA, QMMA) that 
reflect hardware support for low-precision formats. 
Hopper, on the other hand, provides support for \(wgmma\) 
instructions, which enable warp-group asynchronous 
matrix operations, but lacks FP4 and FP6 support. 

\begin{table}[h]
    \scriptsize
    \centering
    \begin{tabular}{|>{\raggedright\arraybackslash}p{0.15\linewidth}|>{\centering\arraybackslash}p{0.35\linewidth}|>{\centering\arraybackslash}p{0.35\linewidth}|}\hline
         &\textbf{GB203 (5th-Gen)}& \textbf{GH100 (4th-Gen)}\\\hline
         Supported Datatypes&FP4, FP6, FP8, INT8, FP16, BF16, TF32, FP64&FP8, INT8, FP16, BF16, TF32, FP64\\\hline
         MMA Instructions&mma, wmma, tcgen05&mma, wmma, wgmma\\\hline
    \end{tabular}
    \vspace{.2cm}
    \caption{Tensor core supports datatypes and \(mma\) instructions. tcgen05 is yet to be supported for the arch. sm\_120a.}
    \label{tab:tensor-core-data-types}
    \vspace{-0.5cm}
\end{table}

\subsection{Variable MMA and Tile-Based Instructions}

Even though the \(wgmma\) instruction isn't supported in 
\blackwell{} and the \(tcgen05\) instruction hasn't 
been implemented yet for \blackwell{}, NVIDIA has 
implementation of the \(mma\) instruction in both 
GPUs with the respective datatypes. 
instruction for the tensor cores to be analyzed.

The matrix multiplication accumulate (MMA) operation 
enables matrix computations for GEMM and deep learning 
workloads. Each \(mma\) instruction specifies a tile 
shape, denoted as M$\times$N$\times$K, which determines the dimensions of the matrix fragments processed 
 per warp or per threadgroup. For example, the instruction in Equation \ref{eq:basic-mma} computes a 16$\times$8 (M$\times$N) 
 output tile using 16$\times$32 (M$\times$K) and 
 32$\times$8 (K$\times$N) inputs. 
 \begin{equation}\label{eq:basic-mma}
     mma.sync.aligned.m16n8k32.f32.f16.f16.f32
 \end{equation}

There are a variety of other supported tile shapes 
such as \(m8n8k16\) or \(m16n8k64\), that support 
finer granularity or larger operand reuse per 
instruction issue. Adjacent to tile shapes, \(mma\) 
instructions support various input/output precisions 
including but not limited to FP4, FP8, FP16 and FP32. 
These datatypes are encoded in Eq. \ref{eq:basic-mma}, 
 where f16 and f32, denotes FP16 inputs with 
 FP32 accumulation and outputs. 
 
In the new instruction set released for Blackwell, 
CUDA 12.9, the \textit{.kind::f8f6f4} suffix must be 
explicitly specified on the PTX instruction to use 
FP6 or FP4 \(mma\) operations on \blackwell{}. 
Attempts to use these formats on \textbf{GH100} or 
without the \(kind\) specifier result in PTX errors. 
Table \ref{tab:dt_comparison} shows the matrix shapes 
and PTX instructions after 
\textit{mma.sync.aligned.kind::f8f6f4} used 
across precision formats, that were tested in our 
microbenchmarks.

\begin{table}
    \centering
    \begin{tabular}{|c|l|l|}\hline
         \textbf{Format}&\textbf{D-Types}&\textbf{PTX Instruction}\\\hline
         e2m1&FP4 &.m16n8k32.row.col.f32.e2m1.e2m1.f32\\\hline
 e3m2& FP6&.m16n8k32.row.col.f32.e3m2.e3m2.f32\\\hline
         e2m3&FP6 &.m16n8k32.row.col.f32.e2m3.e2m3.f32\\\hline
 e4m3& FP8&.m16n8k32.row.col.f32.e4m3.e4m3.f32\\\hline
 e5m2& FP8&.m16n8k32.row.col.f32.e5m2.e5m2.f32\\\hline
    \end{tabular}
    \vspace{0.2cm}
    \caption{Comparison of the supported datatypes (D-Types) on the 4th and 5th generation tensor cores that are being tested with the mma instruction. e8m0 is only used for scaling exponents in the block so it was not tested\cite{NVIDIA_PTX_8.8}. }
    \label{tab:dt_comparison}
    \vspace{-0.5cm}
\end{table}

Through our experiments the PTX-level \(mma.sync\) 
instructions are translated into, OMMA, QMMA, or HMMA 
SASS instructions. By observing the generated SASS 
instructions on the \textbf{GH100}, we observe each \(mma.sync\) 
uses the HMMA instruction for each datatype. 
For Blackwell, the CUDA Binary Utilities 12.9 
documentation \cite{nvidia_binary_utils} specify 
QMMA is used for FP8 matrix multiply and accumulate 
across a warp, while OMMA is used for FP4 matrix 
multiply and accumulate across a warp. Our 
microbenchmarks confirmed both formats of FP8 
inputs use the new QMMA instruction as well as 
both formats for FP6 inputs. While the FP4 input 
\(mma\) was intended to use the OMMA SASS instruction, 
instead the QMMA instruction was observed. However, 
when using block scaling with FP8 \(ue8m0\) as the 
scaling format, OMMA was observed in the SASS code. 
Suggesting that QMMA is the fall back for FP4 inputs 
in the current software.

\textit{In summary, with limited software support as NVIDIA is starting to develop these features, we want to provide a current understanding of the usability and a deep analysis of these pipelines. 
}

\subsection{Precision Tradeoffs}

Low-precision formats are used to reduce memory 
footprint and improve throughput, especially for 
inference workloads. The supported datatypes mentioned 
in the previous section (FP4, FP6, FP8) all have different 
formats and are considered low-precision. These datatype
 formats reduce the number of bits used to represent 
 floating-point numbers by adjusting the number of 
 exponent and mantissa bits, hence a trade-off in 
 dynamic range and precision. 

Previous studies have worked to understand the 
accuracy of these low-precision formats, in this 
section we will analysis the performance-per-watt 
and power consumption of these low-precision formats 
on both architectures. 

\begin{table}
    \centering
    \begin{tabular}{|c|l|l|}\hline
         Data Formats&Blackwell& Hopper  \\\hline
         FP4 e2m1&  16.753 &n/a\\\hline
 FP6 e2m3& 39.383&n/a\\\hline
         FP6 e3m2&  46.723& n/a\\\hline
 FP8 e4m3& 46.661&55.823\\\hline
         FP8 e5m2&   46.806&  55.786\\\hline
    \end{tabular}
    \vspace{0.2cm}
    \caption{Power Usage (Watts)/Performance per Watt  with data formats on the Blackwell and Hopper architectures. }
    \label{tab:low_precision}
    \vspace{-0.5cm}
\end{table}

Noticeably, Table \ref{tab:low_precision} summarizes 
our microbenchmark results on \blackwell{} and \textbf{GH100}.
 \textbf{GH100} lacks native support for FP4 and FP6 
 formats but sustains slightly higher power usage with 
 both FP8 formats (roughly 55W), while \blackwell{} 
 largest power consumption is 46W for the same formats. 
 Power usage generally decreases with lower precision 
 as FP4 achieves the lowest consumption at 16.75W, 
 while FP6 and FP8 formats draw over 39W and 46W, 
 respectively. 

\textit{These results suggest Blackwell's architectural efficiency at low precision and demonstrate a trade-off between numerical expressiveness and energy consumption in \(mma\) tensor core workloads. 
}
\subsection{Warp Scaling and Shared Memory Access}

With our implemented low-precision input \(mma\) 
microbenchmarks, we vary ILP and warp count to 
inspect instruction mapping and compare warp 
scheduling behavior of \blackwell{} and \textbf{GH100}. 

The maximum ILP level at which sustained throughput 
is achieved for each precision format, across decreasing
 warp counts is ILP=5 with 29 active warps, for 
 \textbf{GH100}, and ILP=6 at 25 active warps, 
 for \blackwell{}. This implies Blackwell is capable 
 of issuing more independent \(mma\) instructions per 
 thread, compared to the lower ILP scaling in the 
 \textbf{GH100}. 

When ILP=1 and warps=1, the cycle count from instruction 
issue to data being usable is the completion latency, 
for \blackwell{} all precision formats is 1.21094 
cycles. \textbf{GH100} completion latency is 1.65625 
cycles, suggesting all low-precision formats for 
\(mma\) instructions use the same execution pipeline, 
on there respective architectures. Similarly, 
\blackwell{} achieves higher throughput than 
\textbf{GH100} across all low-precision formats, 
peaking at over 11 TFLOP/s with 6 ILP and 32 active 
warps. Suggesting that increasing ILP significantly 
boosts throughput at low warp counts, confirming that 
Blackwell's warp scheduler efficiently exploits 
intra-warp parallelism when concurrency is limited.

\begin{figure}
    \centering
    \includegraphics[width=0.8\linewidth]{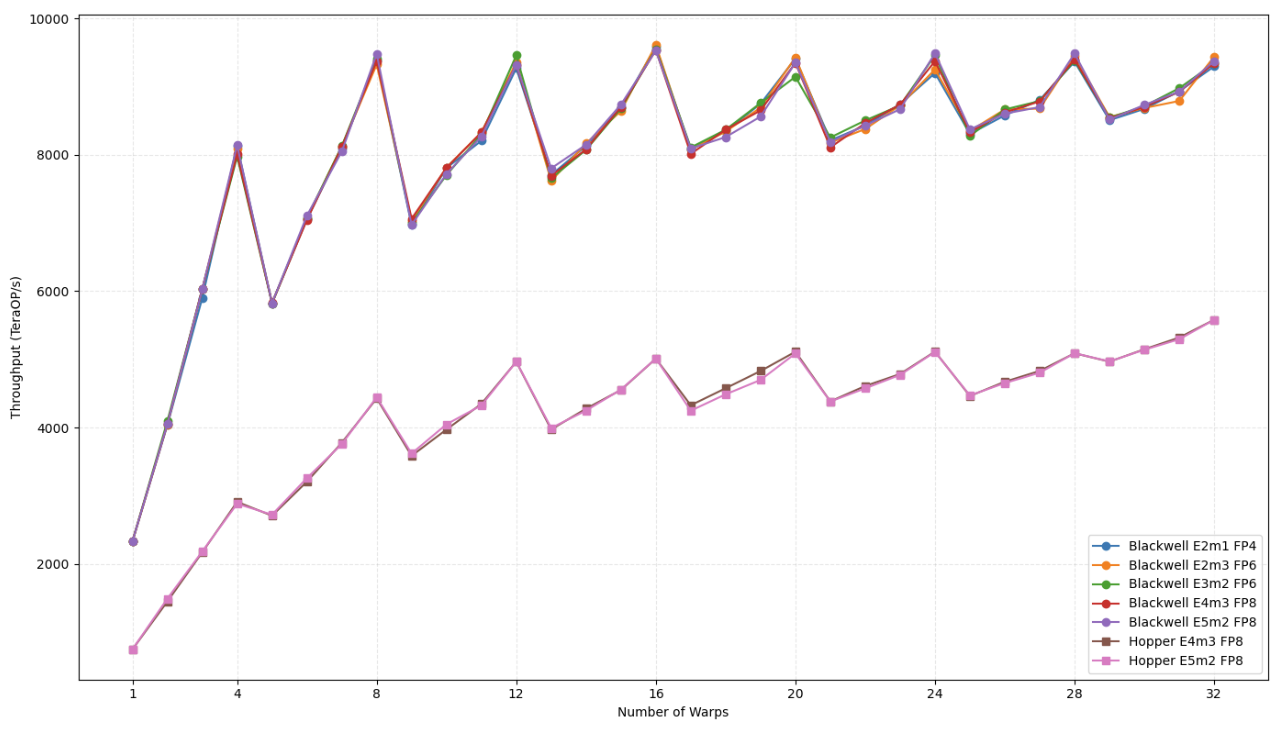}
    \caption{Throughput of \blackwell{} and \textbf{GH100} with varying precision formats and warp counts.}
    \label{fig:tensor-warp-throughput}
    \vspace{-0.3cm}
\end{figure}

We averaged the ILP latency and throughput for each 
low-precision format to present a trend of 
Blackwell and Hopper. As shown in Figure 
\ref{fig:tensor-warp-throughput}, \blackwell{} 
has improved throughput for every precision format. 

\begin{figure}
    \centering
    \includegraphics[width=0.8\linewidth]{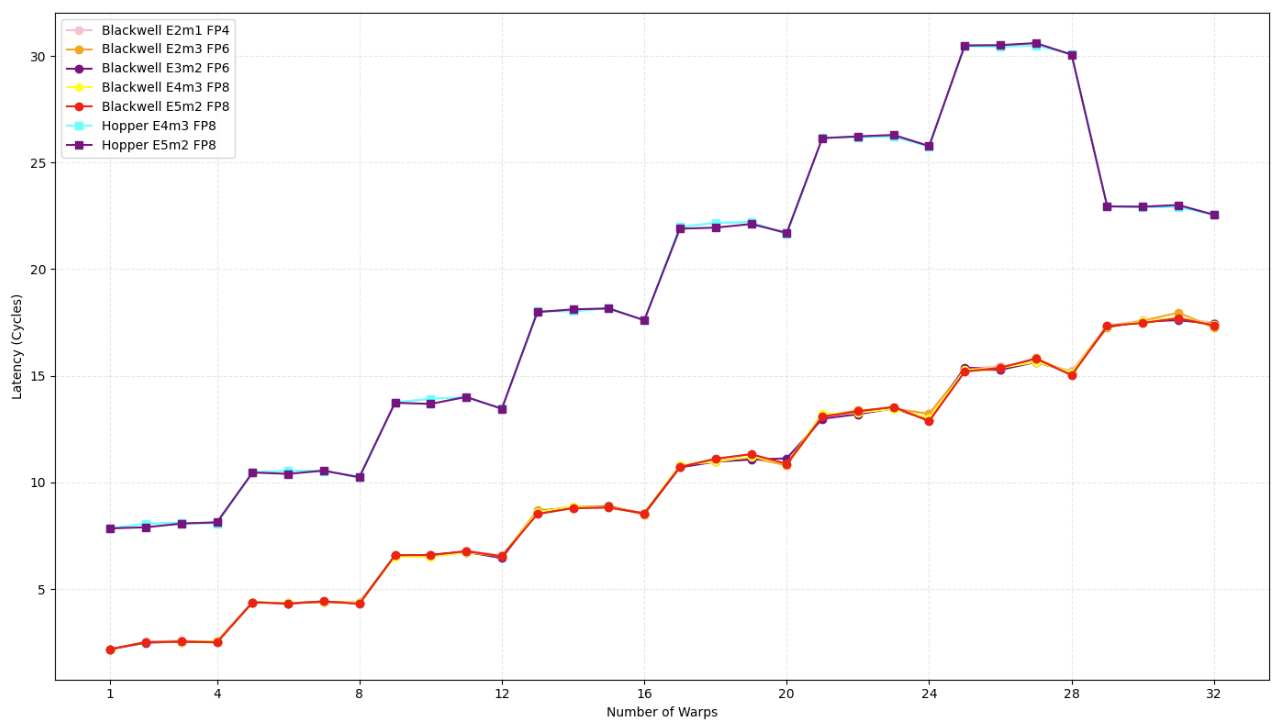}
    \caption{Latency of the \blackwell{} and \textbf{GH100} with varying precision formats and warp counts.}
    \label{fig:tensor-warp-latency}
    \vspace{-0.5cm}
\end{figure}

Similarly, Figure \ref{fig:tensor-warp-latency} 
illustrates latency scaling across formats. 
\blackwell{} sustains consistently lower latency, 
especially for FP4 and FP6, while \textbf{GH100} 
experiences step-like increases in latency as more 
warps are added, a sign of deeper but less agile 
scheduling queues. This indicates that \textbf{GH100} 
requires more warps in flight to saturate execution 
units, whereas \blackwell{} performs better with 
fewer warps but more independent instructions.
Together, these results show that Blackwell’s warp 
scheduler is optimized for low-precision, high-ILP 
workloads with clean control flow, while Hopper relies 
on bulk concurrency and deeper buffering to maintain 
performance under less regular conditions.

This comparison indicates that Blackwell is optimized 
for higher per-thread instruction throughput, while 
both have a similar warp scheduling capacity 
regardless of data formats, reflecting different 
trade-offs in their tensor core microarchitectures. 

\textit{Our methodology can serve as a reference framework for evaluating tensor core performance on future architectures and highlights critical tradeoffs in precision, throughput, and execution behavior at the warp level.
}

\section{Memory Subsystem}
\label{sec:memory}

GPU performance is increasingly constrained by memory subsystem behavior rather than raw computation throughput. Efficient utilization of the memory hierarchy, including shared memory, various levels of cache, and global memory, is crucial for achieving architectural efficiency. While both the \textbf{GH100} and \blackwell{} adopts similar memory layouts, they exhibit distinct trade-offs in latency, bandwidth, and capacity. 

This section presents a comparative evaluation of the memory subsystems through microbenchmarking methodologies that measure latency, saturation behavior, and sensitivity to access patterns.

\subsection{Memory Hierarchy Overview}

This study focuses on device-level memory access, excluding host-device transfer performance, which is heavily influenced by system interconnects (i.e. PCIe vs. NVLink). GPU memory access patterns target global memory, shared memory, and hardware-managed cache layers (L2, L1, L0 i-cache), in addition to the register file. 

\begin{figure}[]
    \centering
    \includegraphics[width=0.8\linewidth]{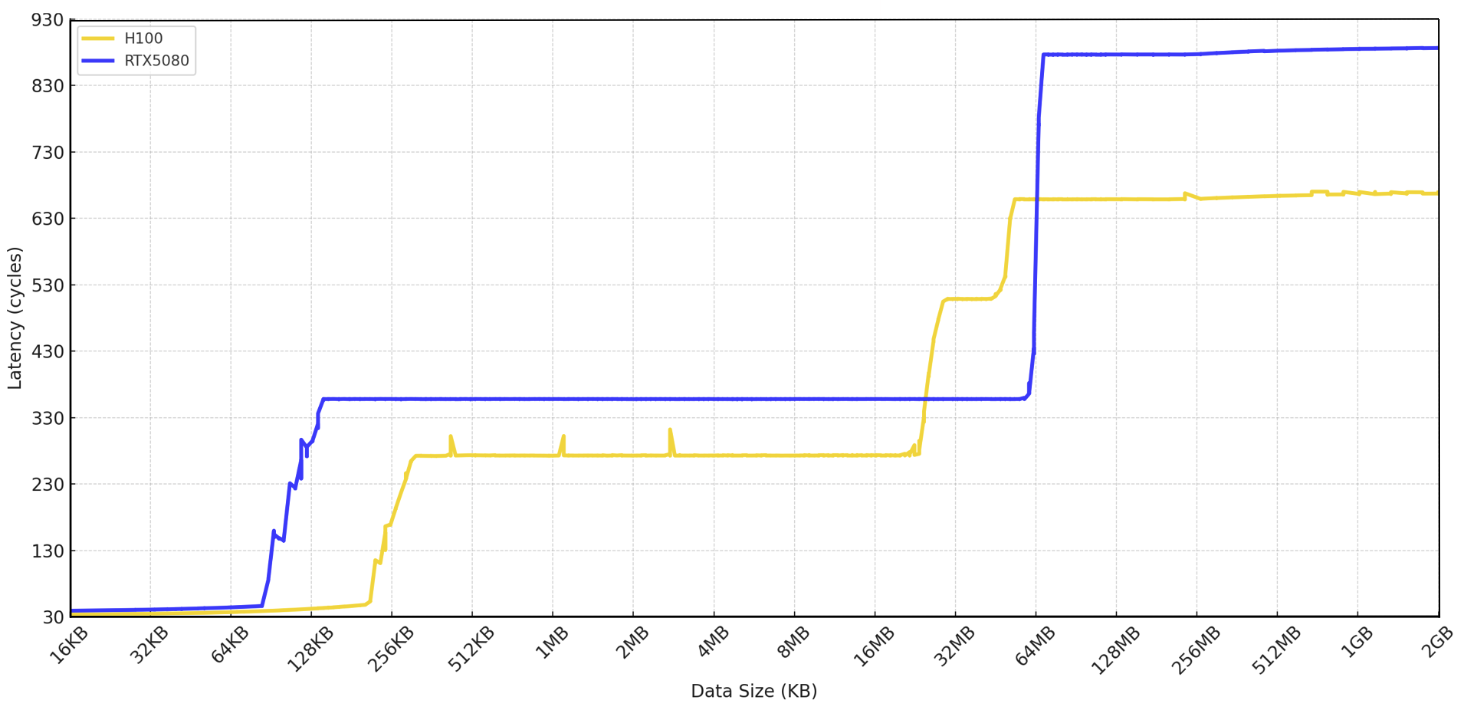}
    \caption{Latency in cycles of the memory hierarchy on the \blackwell{} and \textbf{GH100}.}
    \label{fig:h100-vs-rtx-latency}
    \vspace{-0.5cm}
\end{figure}

To isolate latency characteristics, we employ a pointer-chase microbenchmark with random serialized memory accesses. Figure \ref{fig:h100-vs-rtx-latency} illustrates latency (in cycles) across increasing data sizes for \blackwell{} and \textbf{GH100}.



Three cache regions:

\begin{enumerate}
    \item \textbf{L1 Cache:} spanning 0 to ($\approx$128KB or $\approx$256KB)
    \item \textbf{L2 Cache:} spanning end of L1 to ($\approx$30 MB or $\approx$60MB)
    \item \textbf{Global Memory:} beyond L2 Cache
\end{enumerate}

Latency spikes correspond to cache boundaries, consistent with architectural specifications (Table \ref{tab:cache-hierarchy-details}).

\subsection{Shared Memory and L1 Cache Behavior}

Modern NVIDIA GPUs combine Shared Memory and L1 Cache in a unified memory space per SM. To evaluate the performance and characteristics of this unified design on the \blackwell{} and \textbf{GH100} chips, we developed microbenchmarks to measure access latency trends, bank conflict sensitivity, and warp scaling behavior.

As seen from the pointer-chasing benchmark in Figure \ref{fig:h100-vs-rtx-latency}, both GPUs perform nearly identical latencies in the L1 cache region, reaming consistent at 30-40 cycles, indicating similar hit latencies in the hardware-managed data path. Despite architectural differences, this suggests a well-optimized L1 access path. 

However, cache capacity differs significantly. \textbf{GH100} features up to 256 KB of combined L1/shared memory per SM, whereas \blackwell{} reduces this to 128 KB/SM.

The \textbf{GH100} and \blackwell{} expose a configurable portion of this memory to software as shared memory. Using dynamic allocation from the \(cudaFuncSetAttribute\) and \(cudaFuncAttributeMaxDynamicSharedMemorySize\) attribute, we determined the configurable shared memory limits to be  $\approx$227 KB/SM on \textbf{GH100} and $\approx$99 KB/SM on \blackwell{}. Without dynamic allocation, the default static shared memory limits remains 48 KB/SM on both architectures. 

To explore access behavior, we designed two microbenchmarks. For shared memory, we accessed a statically declared \(\_\_shared\_\_\) array with configurable stride and warp counts. Similarly for the L1 cache, we accessed global memory via a \(\_\_restrict\_\_\) \(float*\) \(gmem\), with working sets designed to fit within L1 cache capacity and induce conflict via strided loads. Both benchmarks swept from 1 to 32 warps and stride sizes of 1 and 4 with 32 memory accesses. Each test was repeated 1024 times, and median latency was recorded.

\begin{figure}
    \centering
    \includegraphics[width=0.8\linewidth]{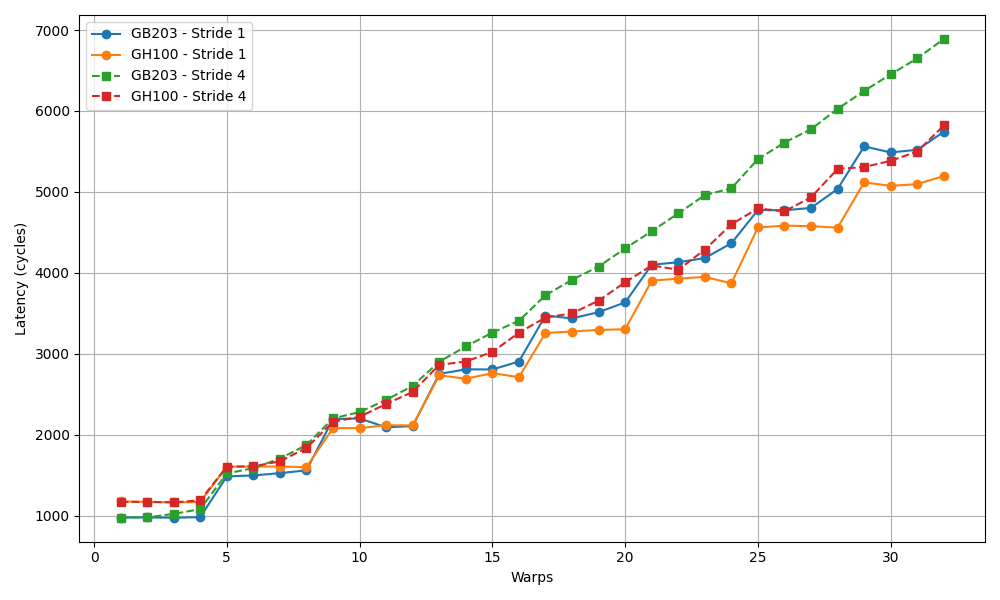}
    \caption{Latency comparison of \textbf{GH100} and \blackwell{} with Shared Memory.}
    \label{fig:bankconflict_latency}
    \vspace{-0.5cm}
\end{figure}

Figure \ref{fig:bankconflict_latency} shows how shared memory latency scales with increasing warp count. For both strides, \blackwell{} exhibited lower latency at low warp counts (1-4 warps), suggesting a more optimized path under light loads. However, \textbf{GH100} outperforms under higher warp pressure (6–32 warps), likely due to its larger shared memory capacity but could be from a more robust bank conflict mitigation. 

With stride 4, \textbf{GH100} maintains smoother scaling and lower latency at a higher warp level, indicating better tolerance to access skew. In contract, \blackwell{} exhibits steeper latency increased with stride 4, likely due to bank contention and saturation of its smaller memory partition. 

\begin{figure}
    \centering
    \includegraphics[width=0.8\linewidth]{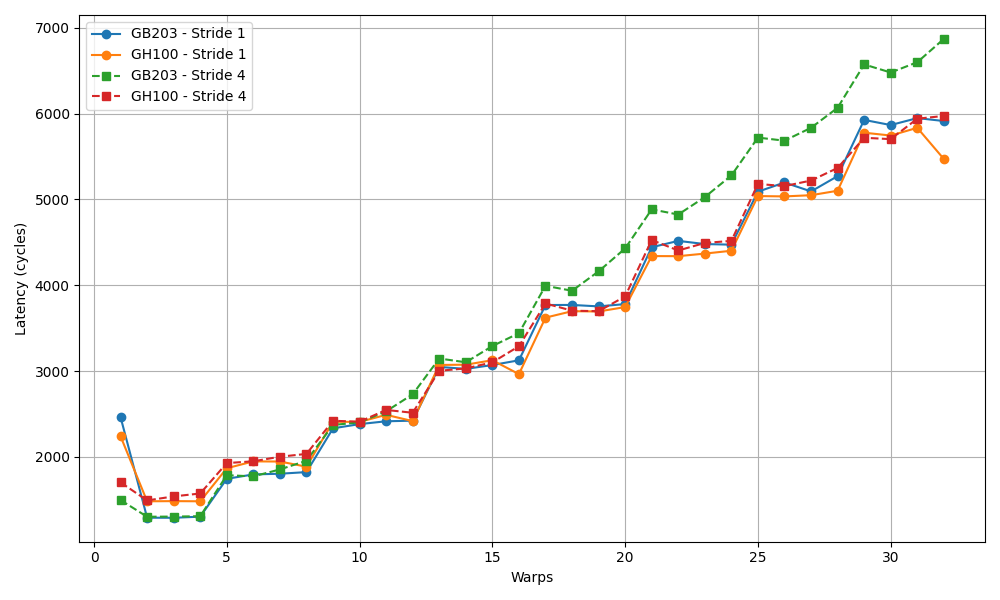}
    \caption{Latency comparison of \textbf{GH100} and \blackwell{} with L1 Cache.}
    \label{fig:l1_latency}
    \vspace{-0.5cm}
\end{figure}

Figure \ref{fig:l1_latency} shows latency trends as more warps access the L1 cache. While both architectures show a latency increase at the first access, \blackwell{} maintains slightly lower latency from 2-11 warps on stride 1. L1 latency remains flatter across warp count steps compared to shared memory, particularly on \textbf{GH100}, likely due to spatial locality and L1's higher tolerance to conflict. However, under stride 4 on \blackwell{}, latency increases more sharply, suggesting that even though L1 is less sensitive to stride than shared memory, access skew still impacts performance, especially with a lower memory partition that can become easily saturated. 

Overall, shared memory latency is highly sensitive to warp count and access stride, particularly on \blackwell{} where bank conflicts scale more aggressively. In contrast, L1 cache exhibits more latency with better resilience as warps increase through shared memory and L1 cache latency meet at 32 warps. \textbf{GH100}'s larger unified memory and smoother warp scaling give it an advantage in highly threaded kernels with dense reuse. \blackwell{}, on the other hand, improves low latency access paths and conflict resolution at small warp counts, likely via microarchitectural enhancements such as multiported banks or warp aware scheduling. 

\textit{These results demonstrate the improvement in warp scaling kernel design, particularly for the \blackwell{}, though the chip is still constrained by the memory partition limits. 
}
\subsection{L2 Cache}

The L2 cache architectures in the \textbf{GH100} and \blackwell{} GPUs reflect two distinct designs, particularly in how they handle partitioning and scaling under load. Positioned between global memory and the SMs, the L2 cache is the largest on-chip memory block. 

In the \textbf{GH100}, the L2 cache is divided into two independent partitions, each servicing a subset of GPCs. This partitioned design supports better data locality and parallelism across cache accesses. In contrast, the \blackwell{} employs a monolithic L2 cache shared by all GPCs. This unified approach simplifies global memory routing and coherence, and can improve spatial locality for smaller or graphics-oriented workloads. However, it may also lead to greater contention when many SMs issue simultaneous uncached or streaming memory accesses.

Latency measurements highlight these differences. For standard L2 hits, the \blackwell{} exhibits a fixed latency of approximately 358 cycles, while the \textbf{GH100} achieves a lower latency of around 273 cycles. This latency advantage in \textbf{GH100} likely stems from its partitioned design, which reduces contention by distributing access across two units. As memory demands grow, however, \textbf{GH100}'s advantage diminishes: when both partitions are saturated, latency increases to about 508 cycles for memory sizes ranging from 31 MB to 45 MB. In contrast, \blackwell{} maintains its baseline latency further into the memory footprint, due to its larger total L2 capacity (65 MB compared to \textbf{GH100}'s 50 MB).

To understand how these architectures perform under warp-level concurrency, we developed a microbenchmark that issues 1024 global memory load/store operations per thread and tracks per-warp cycle timing. This setup allows us to evaluate how L2 throughput scales with increasing warp counts.

\begin{figure}
    \centering
    \includegraphics[width=0.8\linewidth]{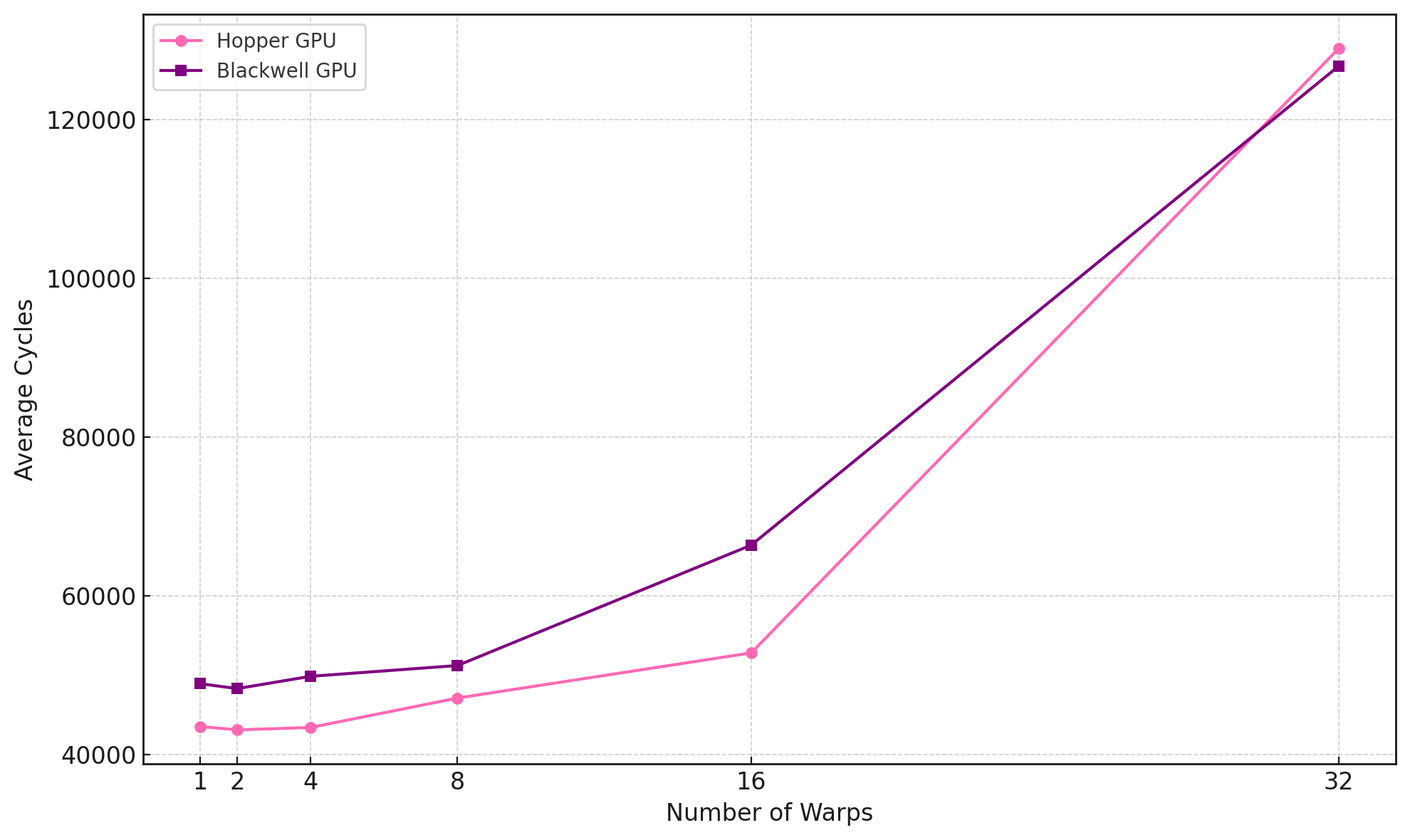}
    \caption{L2 cache latency with warp scaling.}
    \label{fig:l2-warp}
    \vspace{-0.5cm}
\end{figure}

Figure \ref{fig:l2-warp} shows at low warp counts (1–4), the \textbf{GH100} consistently delivers better performance, with average per-warp cycle times around 43.5k, compared to \blackwell{}’s 49k. This difference reflects not only \textbf{GH100}'s faster L2 latency but also its deeper warp scheduler pipeline and more efficient buffering. In the 8–16 warp range, \textbf{GH100} maintains its advantage with minimal performance degradation, whereas \blackwell{} begins to show signs of saturation, reaching approximately 66k cycles at 16 warps. This suggests that \blackwell{}’s single L2 interface becomes a bottleneck as concurrent memory pressure grows.

Interestingly, at high warp counts (16–32), \blackwell{} catches up and eventually slightly outperforms \textbf{GH100} at 20 warps. At 32 warps, \blackwell{} completes the benchmark in $\approx$128.4k cycles per warp, compared to \textbf{GH100}’s $\approx$128.9k. This shift reflects \blackwell{}’s higher aggregate L2 bandwidth under extreme load, likely a result of its larger cache size and reduced partitioning overhead. While \textbf{GH100} offers consistent performance and deterministic warp scheduling at low to moderate concurrency levels, it reaches a throughput ceiling under full pressure, constrained by its partition arbitration.

These trends suggest that \textbf{GH100} is better suited to latency-sensitive and dynamic workloads that operate under moderate concurrency, thanks to its aggressive warp scheduling and partitioned cache layout. \blackwell{}, on the other hand, delivers superior performance under full utilization, making it more favorable for large-scale, bandwidth-bound applications such as deep learning inference or dense matrix operations.

\textit{In summary, \textbf{GH100}'s partitioned L2 architecture is optimized for high concurrency and compute-heavy server-class workloads. \blackwell{}'s unified L2 design simplifies hardware complexity and favors mixed compute-graphics use cases. These architectural trade-offs must be considered when tuning for specific performance targets in memory-bound kernels, whether prioritizing latency, throughput, or data locality.
}
\subsection{Global Memory}

\begin{figure}
    \centering
    \includegraphics[width=0.8\linewidth]{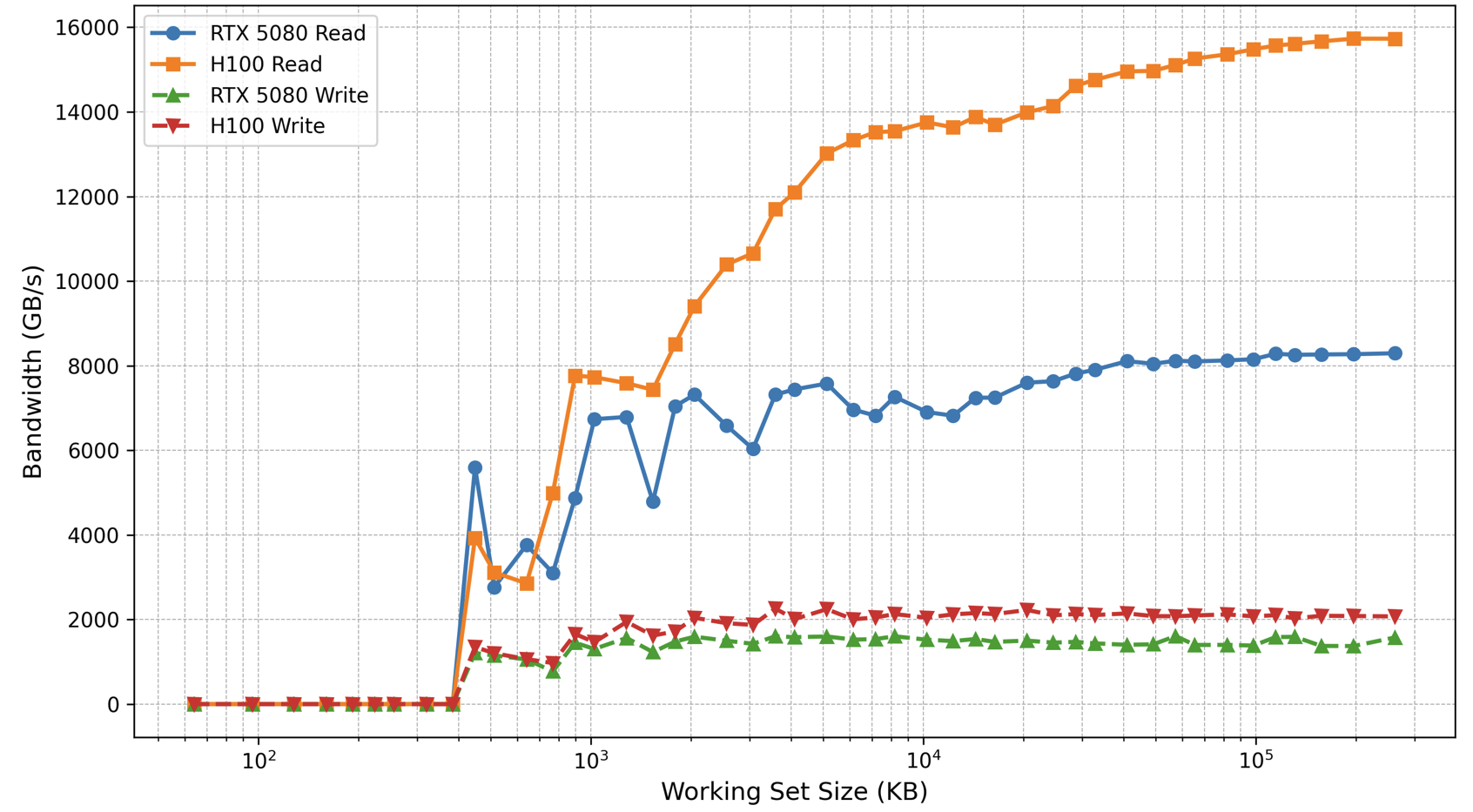}
    \caption{\blackwell{} and \textbf{GH100} throughput of the memory hierarchy.}
    \label{fig:gb}
    \vspace{-0.5cm}
\end{figure}

We extend the analysis to global memory bandwidth using a series of sustained transfer benchmarks. As shown in Figure \ref{fig:gb}, \textbf{GH100} achieves a peak read bandwidth of 15.8 TB/s, substantially higher than \blackwell{}'s 8.2 TB/s. Write bandwidth is lower on both, 2.2 TB/s (\textbf{GH100}) vs. 1.6 TB/s (\blackwell{}), demonstrating the architectural design toward read-heavy workloads. Possibly due to narrower write-back paths or less aggressive write coalescing. 
Latency trends observed in Figure \ref{fig:h100-vs-rtx-latency} indicate global memory access begins beyond 71 MB (\blackwell{}) and 55 MB (\textbf{GH100}), with respective latencies of $\approx$876.7 cycles and $\approx$658.7 cycles. \textbf{GH100}'s superior latency performance is attributable to its use of HBM2e, which offers higher bandwidth and lower latency than \blackwell{}'s GDDR7. 

While the Blackwell architecture introduces notable enhancements in memory scheduling and subsystem design, these changes may lead to reduced consistency in irregular or latency-sensitive workloads.

\section{Microbenchmark Case Studies}
\label{sec:kernel}

To evaluate how microarchitectural differences between \textbf{GH100} and \blackwell{} affect real-world performance, we implemented a set of representative GPU kernels spanning key application domains. These case studies bridge the gap between synthetic benchmarks and practical performance behavior, allowing us to assess how memory hierarchy, warp scheduling, and tensor cores interact in realistic execution environments.

\subsection{Dense GEMM}

A dense general matrix multiplication (D-GEMM) kernel utilizes nearly every stage of the compute pipeline, from shared memory operand staging, register usage, warp scheduling, to tensor core utilization. We evaluate a D-GEMM kernel with FP8 inputs using NVIDIA's cuBLASLt API and the \(\_\_nv\_fp8\_e4m3\) datatype. Our kernel performs a fused matrix multiplication and accumulation of the form \(D = A^T * B + C\), where A and B are FP8, C is represented as bfloat16, and D is stored in FP8.

The goal was to assess compute throughput and power behavior across varying matrix sizes, providing a comparison not only between the \textbf{GH100} and \blackwell{} but also to the lower-level PTX microbenchmarks in previous sections. 

\begin{figure}
    \centering
    \includegraphics[width=0.9\linewidth]{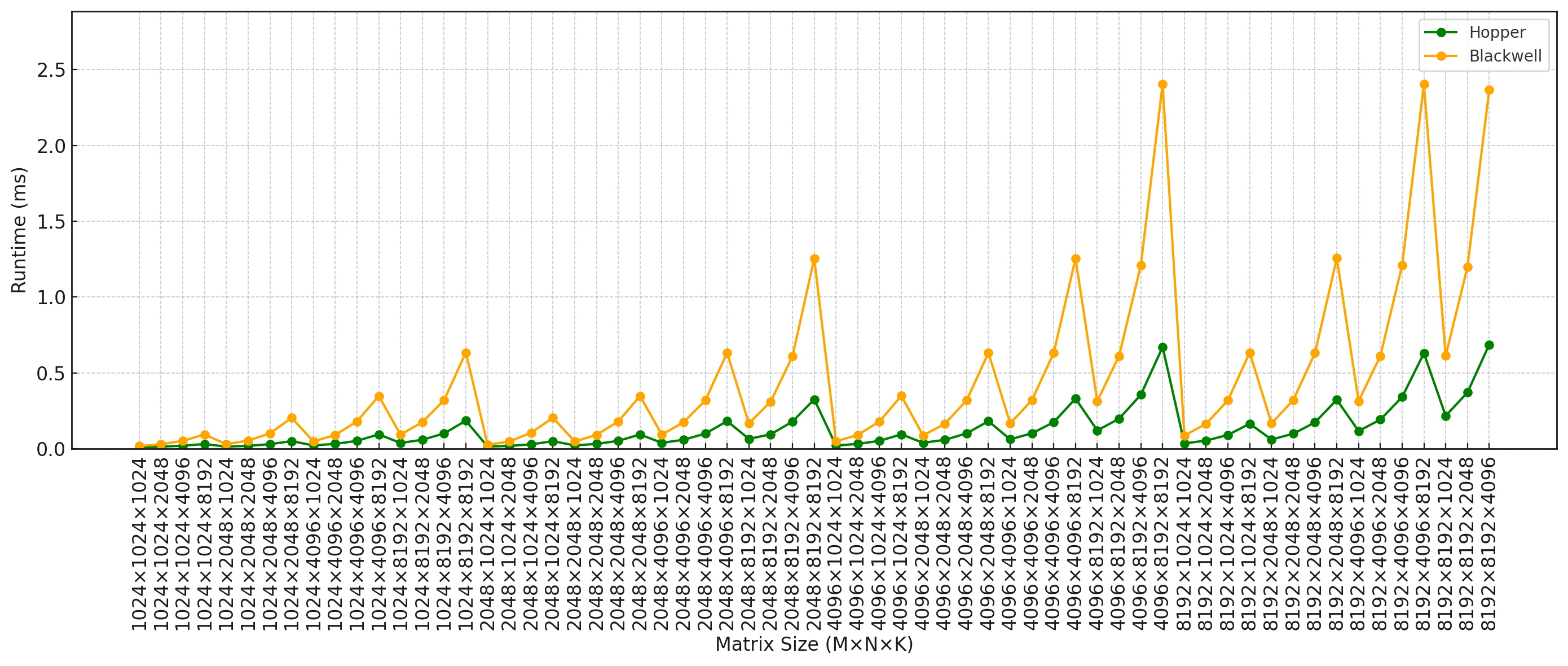}
    \caption{The runtime (ms) of each execution size (M$\times$N$\times$K) on both the Hopper H100 and the Blackwell RTX 5080 GPUs. The M=N=K=8192 kernel runtime was 4.710 ms for the Blackwell GPU, omitted from the graph. }
    \label{fig:gemm-latency}
    \vspace{-0.5cm}
\end{figure}

The GEMM kernel is benchmarked using a 32 MB workspace. We used nvidia-smi to measure power and efficiency. Each configuration was executed 100 times, averaged, and repeated across sizes (1024, 2048, 4096, 8192). Figure \ref{fig:gemm-latency}, shows the runtime comparison across matrix sizes (M$\times$N$\times$K) for both GPUs. Hopper consistently outperforms Blackwell, achieving lower runtime across nearly all configurations. The performance gap widens with larger matrix sizes, where Blackwell shows significant spikes in latency. This inconsistency suggests potential instability in kernel selection or scheduling for FP8 GEMM on Blackwell, despite theoretical improvements for FP8. Hopper benefits from a more mature compiler heuristic or stable scheduling at scale. 
With the increase in latency on the Blackwell, Table \ref{tab:gemm_throughput} shows the achieved throughput in TFLOP/s, calculated with Equation \ref{eq:tflops}, for selected matrix sizes.
\begin{equation}\label{eq:tflops}
    TFLOPS = (2 \times M \times N \times K)/runtime
\end{equation}


Hopper consistently delivers higher effective throughput across all tested shapes. For example, at the largest configuration (8192$\times$8192$\times$8192), Hopper reaches 0.887 TFLOP/s, nearly 4$\times$ higher than Blackwell's 0.233 TFLOP/s. Even at smaller sizes such as 1024$\times$1024$\times$1024, Hopper maintains a clear advantage. While Blackwell theoretically supports higher FP8 compute rates, these results suggest that kernel selection, memory hierarchy utilization, or scheduling are limiting practical throughput on the current RTX 5080. 

\begin{table}[htbp]
\centering
\begin{tabular}{|c|c|c|}
\hline
\textbf{Matrix Size} & \textbf{Hopper (TFLOP/s)} & \textbf{Blackwell (TFLOP/s)} \\
\hline
8192$\times$8192$\times$8192 & 0.887 & 0.233 \\
\hline
2048$\times$2048$\times$2048 & 0.554 & 0.191 \\
\hline
2048$\times$2048$\times$4096 & 0.674 & 0.192 \\
\hline
2048$\times$4096$\times$8192 & 0.759 & 0.217 \\
\hline
1024$\times$1024$\times$1024 & 0.239 & 0.134 \\
\hline
\end{tabular}
\vspace{0.2cm}
\caption{D-GEMM Throughput on \textbf{GH100} and \blackwell{} GPUs}
\label{tab:gemm_throughput}
\vspace{-0.5cm}
\end{table}

\begin{figure}
    \centering
    \includegraphics[width=0.9\linewidth]{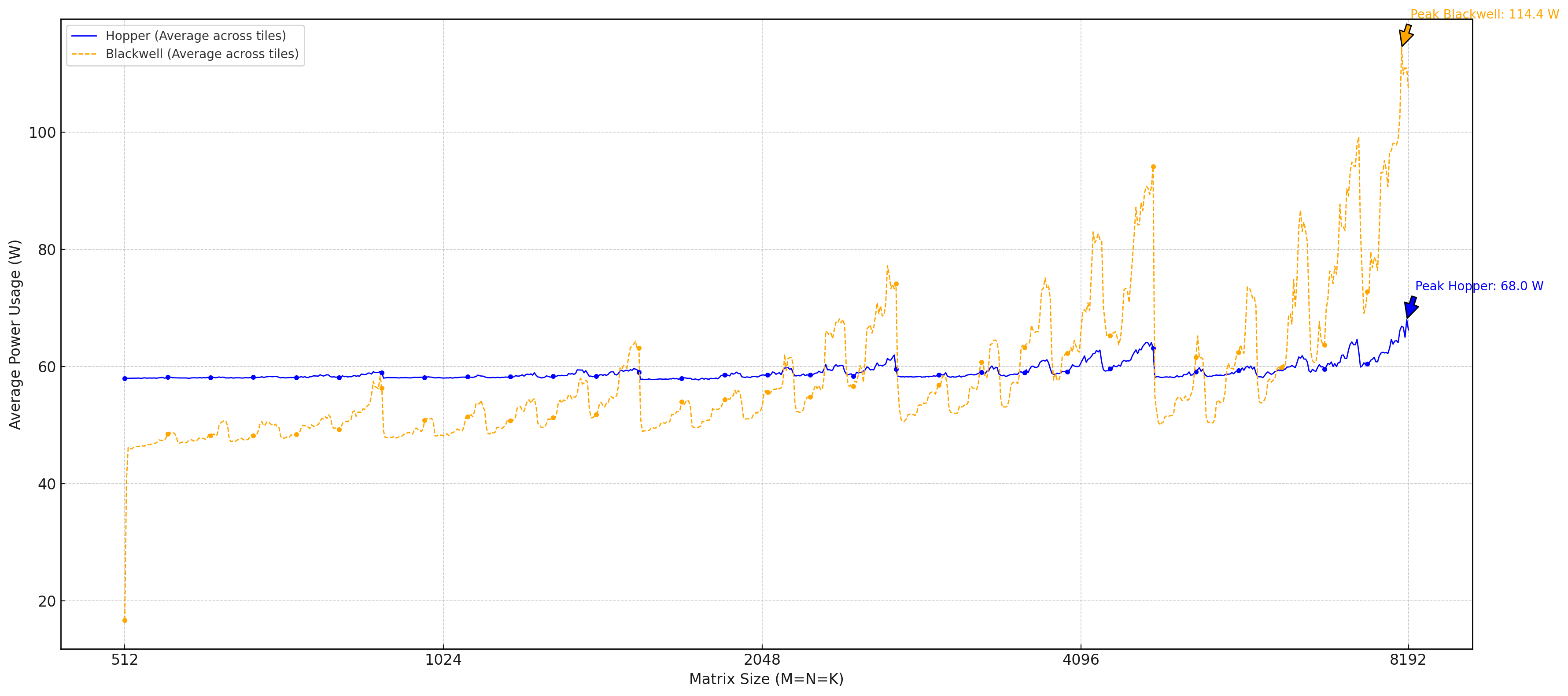}
    \caption{The power consumption (W) of the program with each execution size (M=N=K) on both the Hopper H100 and the Blackwell RTX 5080 GPUs.}
    \label{fig:gemm-power}
\vspace{-0.5cm}
\end{figure}

Lastly we tested tile sizes from 1 to 64 and added a 512 size to the matrix, to see if there was a difference with power since Blackwell had a longer runtime with larger sizes. Figure \ref{fig:gemm-power} compares the average power draw across the matrix sizes. We average the power usage of each tile size at the variable matrix shapes, since the power consumption at each tile size was relatively similar. We still see the overall trend in power consumption with the different matrix shapes. Hopper maintains a relatively flat power profile, with power usage around 58-60W and peaks at 68W even for the largest 8192$\times$8192$\times$8192 matrix shape. On the other side, Blackwell shows higher variability and a stepper power curve, with average power exceeding 80W  and peaking at 114.4W. Notably, there are spikes in power usage when N=K=8192 is used in conjunction with the other sizes. Also for the 512$\times$512$\times$512 matrix shape Blackwell used way less power than compared to Hopper, suggesting Blackwell was able to preserve power during the pipeline, though this was not seen with the rest of the matrix sizes. The higher power of \blackwell{}, combined with lower throughput, results in lower performance-per-watt across most configurations. 

Overall, while Blackwell's RTX 5080, FP8 compute is impressive, real-world efficiency on dense GEMM remains more favorable on Hopper with current software and kernel implementations. 

\subsection{Transformer Inference}

To evaluate performance and energy efficiency under real-world inference workloads, we implemented a Transformer inference case study using TensorRT, NVIDIA's optimized framework \cite{nvidia_tensorrt}. This test complements our dense GEMM benchmark by incorporating memory-intensive and latency-sensitive compute patters such as multi-head attention, MLP layers, layer normalization, and token sampling. 

We selected the GPTneox model \cite{black2022gpt} due to its small size and compatibility with both FP8 and FP4 quantization paths. The model was run with variable precision (best, normal, fp16, fp8). TensorRT uses the best precision for performance or the default precision set which seems to be either FP32 or TF32, for best and normal precisions respectively. Each inference was run a hundred times and the metrics were averaged. Table \ref{tab:inference_power} shows that the Blackwell GPU benefits from a better power model. Hopper maintained a 57-60W consistent power usage across precisions, indicating a stable runtime efficiency. With Blackwell having a more pronounced reduction in power as precision decreases, with a from` 58.8W to 45W in FP8, suggesting better scaling or lower utilization under reduced precision. Interestingly, the "Best" configuration, which should reflect the highest performing engine by TensorRT, shows Blackwell increased power draw.

\begin{table}[htbp]
\centering
\begin{tabular}{|c|c|c|}
\hline
\textbf{Precision} & \textbf{Hopper} & \textbf{Blackwell} \\
\hline
FP32 & 60.24 & 58.82 \\
\hline
FP16 & 57.64 & 47.78 \\
\hline
FP8 & 57.69 & 45.14 \\
\hline
Best & 60.15 & 61.03 \\
\hline
\end{tabular}
\vspace{0.2cm}
\caption{Average Inference Power Consumption in watts Across Precision Models.}
\label{tab:inference_power}
\vspace{-0.5cm}
\end{table}

\textit{Overall, this demonstrates that Hopper delivers a steadier power efficiency across formats, Blackwell can be tuned for competitive inference workloads with better power efficiency. }
\vspace{-0.5cm}

\section{Conclusion}
This work presented a detailed experimental analysis of NVIDIA's Blackwell architecture (\blackwell{} chip) through carefully designed microbenchmarks. By comparing microarchitectural features against the Hopper (\textbf{GH100} chip) GPU, we provide insights into Blackwell's advancements in memory hierarchy, SM execution pipeline, and its 5th-gen Tensor Cores. Our analysis highlights the hardware's increased support for low-precision formats such as FP4 and FP6, revealing their practical implications for power and performance efficiency. The guidelines and observations presented in this study provide a microarchitectural understanding to assist developers in optimizing software to effectively use the hardware and thus enable more efficient deployment of AI and HPC workloads.

\section*{Acknowledgment}

We thank Nikhil Jain from NVIDIA for their correspondence and for addressing our questions. This research used resources on the Frank cluster at the University of Oregon. This material is based upon work supported by the U.S. DOE under Contract DE-FOA-0003177, S4PST: Next Generation Science Software Technologies Project.

\bibliographystyle{IEEEtran}
\bibliography{references}

\end{document}